\documentclass{article}
\usepackage[utf8]{inputenc}
\usepackage[T1]{fontenc}
\usepackage{amsmath}
\usepackage{amssymb}

\usepackage[table]{xcolor}
\usepackage{acronym}
\usepackage{graphicx}
\usepackage{subcaption}
\usepackage[margin=1in]{geometry}

\title{Stopping power monitoring during proton therapy by means of prompt gamma timing: first experimental results with a homogeneous phantom}

\author{Julius~Werner$^{1,\dagger}$ \and
Francesco~Pennazio$^{2,\dagger}$ \and
Piergiorgio~Cerello$^2$ \and
Elisa~Fiorina$^2$ \and
Simona~Giordanengo$^2$ \and
Felix~Mas~Milian$^{2,3,*}$ \and
Alessio~Mereghetti$^4$ \and
Franco~Mostardi$^5$ \and
Marco~Pullia$^4$ \and
Sahar~Ranjbar$^{2,5}$ \and
Roberto~Sacchi$^{2,5}$ \and
Anna~Vignati$^{2,5}$ \and
Magdalena~Rafecas$^{1,\ddagger}$ \and
Veronica~Ferrero$^{2,5,\ddagger}$
}

\date{$^1$Institute of Medical Engineering, Universität zu Lübeck\\
$^2$INFN sezione di Torino\\
$^3$State University of Santa Cruz\\
$^4$Fondazione CNAO\\
$^5$Università degli Studi di Torino\\
$^*$masmilia@to.infn.it\\
$^\dagger$these authors contributed equally to this work\\
$^\ddagger$these authors jointly supervised this work}

\begin{document}
\flushbottom
\maketitle

\thispagestyle{empty}

\begin{abstract}
Proton therapy’s full potential is limited by  uncertainties that prevent optimal dose distribution.
Monitoring techniques can reduce these uncertainties and enable adaptive treatment planning. Spatiotemporal Emission Reconstruction from Prompt-Gamma Timing (SER-PGT) is a promising method that provides insights into both particle range and stopping power, whose calculation would normally require knowledge about patient tissue properties that cannot be directly measured.
We present the first experimental results using a 226.9 MeV synchrotron-proton beam impinging on a homogeneous phantom at a sub-clinical intensity (2 - $4 \times 10^7$ pps). SER-PGT  uses data from a multi-detector setup:  a thin and segmented Low Gain Avalanche Diode for proton detection and Lanthanum Bromide-based crystals for photon detection.
The estimated stopping power profile showed an 8\%\textpm 3\% average error compared to NIST PSTAR values, and 2\%\textpm2\% deviation relative to water at 100 MeV. Range assessment in a phantom with a 4 cm air-gap successfully identified the range shift with a 3 mm standard deviation.
These results demonstrate the feasibility of using SER-PGT to recover both range and stopping power information through particle kinematics and PGT measurements.

\end{abstract}

\section{Introduction} 
The main advantages of Proton Therapy (PT) over conventional radiotherapy include better optimization of the dose delivered to the tumour with respect to the total irradiated dose and higher relative biological effectiveness \cite{Tommasino_2015,Paganetti_2022, Paganetti_2025}. 
Treatment planning generally starts with outlining the Gross Tumour Volume (GTV) and the Organs At Risk (OARs), exploiting data from imaging modalities such as CT. The GTV is then enlarged to account for possible disease spread in the surrounding tissues, forming the Clinical Target Volume (CTV). In photon radiotherapy, the Planning Target Volume (PTV) is then introduced, enlarging the CTV. The PTV is defined to ensure correct dose delivery to the target while mitigating possible setup errors. However, the PTV concept cannot be straightforwardly translated to particle therapy, because it assumes that the shape of the dose distribution remains largely unaltered by the underlying changes of the patient geometry, but this is not granted in the case of protons and heavy ions. Consequently, in particle therapy, the final impact of the uncertainty sources on dose distributions must be carefully considered. Hence, robust optimization methods directly incorporating uncertainties into the planning procedure have been developed \cite{Unkelbach_2018,Paganelli_2024}.

The Treatment Planning System (TPS) designs the plan according to the medical prescriptions and limitations, considering many uncertainty factors \cite{Paganetti_2012,Durante_2016,Graeff_2023}. These include so-called setup errors, which could be mitigated by increasing the precision of the treatment delivery (e.g. improving the beam monitoring detectors) and of the patient positioning. Still, the lack of direct knowledge of some of the patient’s tissue properties, such as the mean excitation energy, further increases uncertainty.

Although not yet in use in clinical routine, treatment monitoring would allow in-vivo verification of the interaction of the irradiated beam with the patient’s tissues. In the short term, it would assure the compliance of the irradiation with the prescription for each treatment fraction. In the long term, treatment monitoring could reduce some of the treatment uncertainties, thus validating some of the tissue property calculations performed by the TPS.

Treatment monitoring can be achieved through the exploitation of secondary radiation arising from patient-beam nuclear interactions. Some techniques have already been tested in patients, such as in-beam PET \cite{Fiedler_2010,Nishio_2010,Ferrero_2018,Tashima_2024}, in-room PET \cite{Bauer_2013}, after-treatment PET \cite{Parodi_2007}, prompt gamma imaging \cite{Richter_2016, Xie_2017} and secondary charged particle tracking \cite{Toppi_2021,Kelleter_2024}. The mentioned techniques exploit the comparison of the measured quantity with a reference, given either by a Monte Carlo simulation or by a measurement of a previous fraction.

The Prompt Gamma Timing (PGT) technique\cite{Golnik_2014} is based on the measurement of the Time Of Flight (TOF) between the primary proton arrival time ($t_0$) and the detection time of a secondary prompt gamma photon ($t_\gamma$). The shape of the PGT measurements distribution depends on several factors, such as particle energy, target composition, and detector position with respect to the target. PGT has been used as an indirect way to assess the particle range in some studies, addressing both experimental \cite{Hueso_Gonz_lez_2015,Marcatili_2020,Jacquet_2023} and theoretical \cite{Jacquet_2021,Schellhammer_2022} aspects. To our knowledge, the results published so far have focused on measuring beam range shifts, and on detecting differences with respect to a reference measurement. The method used in this study builds upon the PGT technique and, by using multiple prompt-photon detectors, leverages the complementary information provided by their varied positions.

Although based on the same measurement principle of the aforementioned techniques, our approach introduces an innovative methodology consisting of two distinct steps: (1) estimating the distribution of the prompt-gamma origins in space and time, and (2) deriving the stopping power from that  distribution.

Spatiotemporal Emission Reconstruction (SER-PGT) had previously been applied only to Monte Carlo simulations, where it was used to recover range shifts \cite{Werner_2024} and, through fits of the 2D space-time distributions with a proton-motion model, we performed a direct calculation of the target stopping power over a broad energy range \cite{Werner_2024, Ferrero_2022a}. 

In this study, we applied the method to experimental data for the first time. We measured PGT distributions from a 226.9 MeV synchrotron proton beam impinging on a homogeneous phantom and processed them with SER-PGT \cite{Pennazio_2022}, to obtain the corresponding 2D space-versus-time distribution of the prompt-photons emission. 
This work presents the first experimental evaluation of the target stopping power and beam range during PT, which is the first step towards the clinical validation of the proposed technique.

\section{Methods}
\subsection{Experimental data}

The experimental measurements were performed at the Centro Nazionale di Adroterapia
Oncologica (CNAO, Pavia, Italy)\cite{Rossi_2022}, with proton beams.

\subsubsection{Multi-Detector PGT-System}
\label{sec:Detectors}

The timestamp of the primary protons was provided by a detector based on a thin and segmented Low Gain Avalanche Diode (LGAD) \cite{Siviero_2022,Sola_2024} sensor. LGADs are n-in-p silicon sensors featuring a moderate internal gain due to a p+ additional layer located below the n++ electrode of a heavily doped junction. The enhanced signal, combined with a small active thickness, leads to time resolutions of about 30 ps for minimum ionizing particles \cite{Siviero_2022}. The sensor, designed for time measurements \cite{Vignati_2023}, was manufactured by Fondazione Bruno Kessler (FBK, Trento, Italy) as part of a production fully dedicated to beam monitoring applications in proton therapy \cite{Marti_Villarreal_2023}. It features a total area of $4.0\times6.6~\text{mm}^2$ segmented in 11 strips of $4.0\times0.54~\text{mm}^2$  area, 0.59~mm pitch and 60~${\mu}$m active thickness over a total thickness of 630 ${\mu}$m. A custom 8-channel analog front-end board was used to read out the eight central strips of the sensor providing signal pulses of less than 2~ns duration with amplitudes of approximately 3~mV per fC of collected charge. Given the effective surface, it was estimated that, with the selected proton beam energy, an average geometrical efficiency of 24\% is achieved.

Prompt photons were detected with two Cerium-doped Lanthanum Bromide (LaBr$_3$:Ce) crystals coupled
with Hamamatsu R13089-11 PhotoMultiplier Tubes (PMTs). LaBr$_3$:Ce was found suitable for
measurements at high count rates (such as the environment found within a clinical proton
irradiation) \cite{L_her_2012} and has already been tested for the specific PGT-application at sub-clinical rates \cite{Hueso_Gonz_lez_2015, Marcatili_2020}. We used two cylindrical crystals (3.81 cm diameter, 3.81 cm height) from EPIC crystals (https://www.epic-crystal.com/), which feature a high photon yield ($6.8\cdot10^4$ photons per MeV) and a decay time of 16 ns. Time and energy resolutions of about 120 ps ($\sigma$) and 7\% $\Delta$E/E, respectively, were found at 1 MeV after characterization with radioactive sources \cite{Ferrero_2022a}.

\begin{figure}[ht]
    \centering
    \begin{subfigure}{0.48\linewidth}
        \centering
        \includegraphics[width=\linewidth]{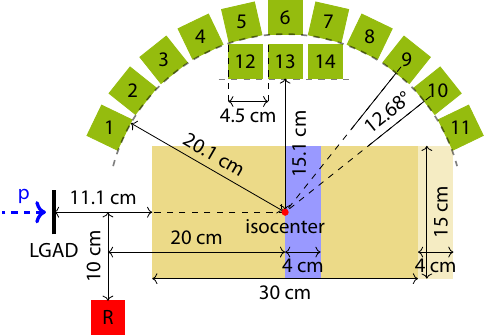}
        \caption{}
        \label{fig:fig1a}
    \end{subfigure}
    \hfill
    \begin{subfigure}{0.48\linewidth}
        \centering
        \includegraphics[width=\linewidth]{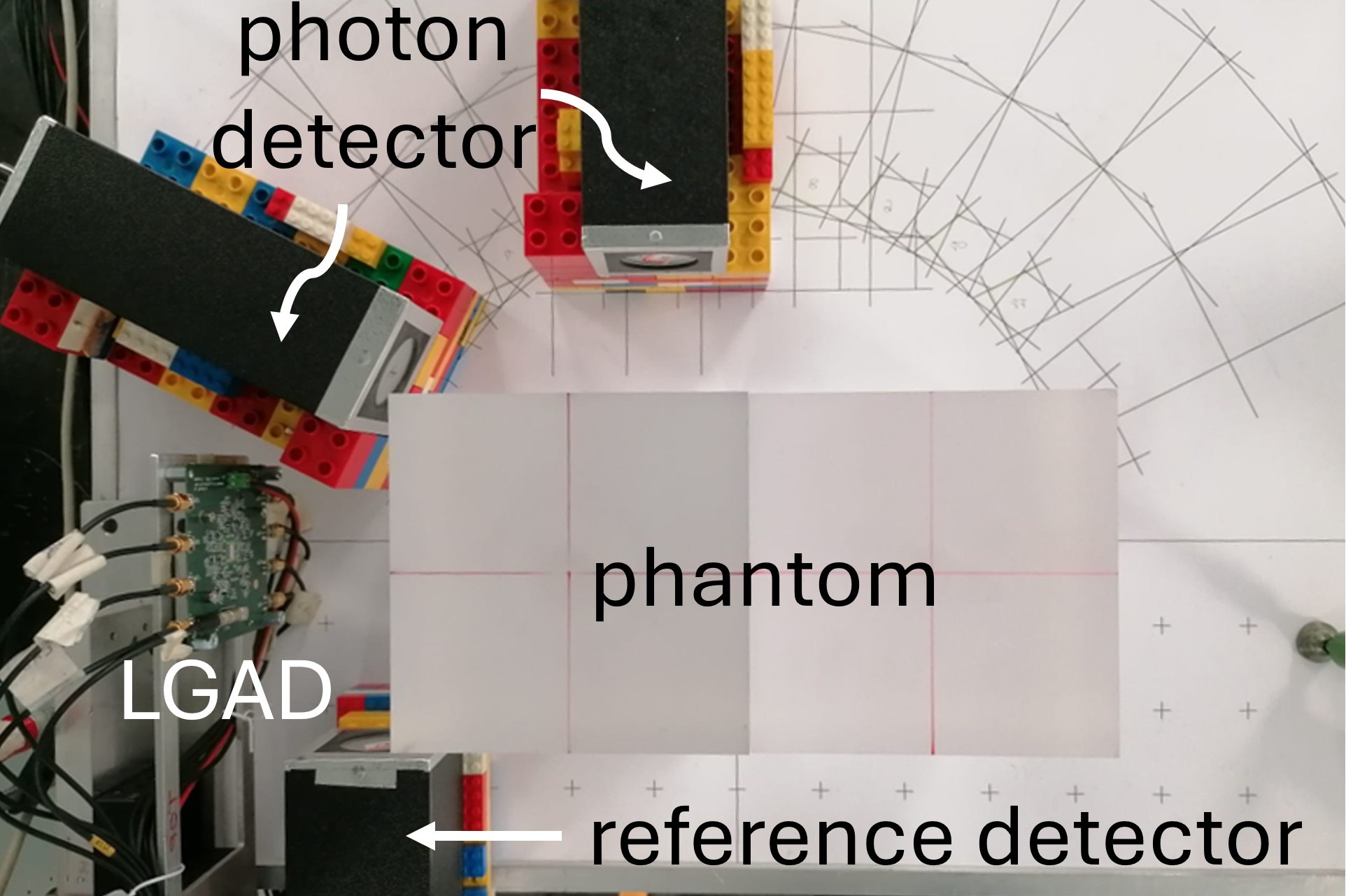}
        \caption{}
        \label{fig:fig1b}
    \end{subfigure}
    \caption{(a): Schematic of the phantom without (yellow) or with (blue) the airgap, the reference detector (R, red), and the photon detector positions (1-14, green). (b): setup in treatment room with the photon detectors in positions 1 and 12.}
    \label{fig:fig1}
\end{figure}

For SER-PGT and, consequently, the stopping power evaluation, a high number of detectors is
optimal \cite{Pennazio_2022, Ferrero_2022a}. However, previous work has shown through Monte Carlo simulations that even a reduced number of detectors can still provide a good estimate of the desired quantities \cite{Ferrero_2022b}. Indeed, a non-symmetrical geometry with detectors placed at different angles and positions yields PGT spectra with distinctive shapes, which can be used for range and stopping power evaluations (see Section~\ref{sec:Range}). For the current study, we selected 14 different positions around the target (Figure~\ref{fig:fig1}). 

For the measurements presented here, only two detectors were available at the same time;
therefore, to acquire the 14 positions, 7 different measurements had to be performed with the same beam/target configuration. At sub-clinical beam intensities, dose delivery may become irregular, resulting in significantly larger fluctuations than those typically observed in clinical settings; consequently, a third detector in a fixed position was used as a reference. This detector, hereafter called the reference detector, was made by coupling a LaBr3:Ce crystal (same characteristics as before) to a $5\times5$ FBK SiPM tile ($10^6$ cells, 15~$\mu$m cell size, $24\times24~\text{mm}^2$ tile size) read out by an in-house developed board \cite{Bartosik_2025}. The reference detector is part of another PGT system presently under development. It can sustain clinical rates, but  currently has a timing resolution of 180 ps $\sigma$. In our case, the PGT distribution of the reference detector was exclusively used to equalize the output of the other detectors in order to merge the 7 measurements (see below).

\subsubsection{Proton beam and target}
\label{sec:Beam}
We selected a proton beam energy of 226.9 MeV, corresponding to a Bragg Peak depth of 31.8 cm in water. Details about the beam shape and other features are described in Mirandola et al. \cite{Mirandola_2015}. The Bragg peak position was verified with a FLUKA Monte Carlo simulation \cite{Ballarini_2024} and found to be  consistent with the expected depth. We also compared the 80\% value of the simulated dose distal fall-off with the projected range of the NIST PSTAR database (https://physics.nist.gov/PhysRefData/Star/Text/PSTAR.html). The two depths are expected to be in very close agreement \cite{Gottschalk_2018}, which is consistent with the $\approx 0.5~\text{mm}$ discrepancy that we found.

At CNAO, the nominal clinical beam intensity (defined as the time-averaged particle fluence rate integrated over the beam's transverse surface) is $2\cdot 10^9$ protons per second (pps). Due to the beam’s bunched structure, the instantaneous rate during each bunch increases by one or two orders of magnitude. Since this clinical intensity exceeds the capacity of our current acquisition system, we conducted our tests at a sub-clinical intensity of approximately $2~ - ~ 4\cdot 10^7$ pps, corresponding to an instantaneous rate about ten times higher than the average value. To adjust the beam intensity, the betatron core that normally pushes the beam into the resonance to drive the slow extraction was switched off. The accelerator was then tuned in such a way that a small part of the beam was sufficiently near to the unstable region to be extracted after the resonance sextupole was turned on. In this way, a halo of the beam was extracted, providing the required lower intensity.

We used a custom experimental setup that samples the beam, measures particle timing, and continuously monitors both the average and instantaneous particle rates, as described in Section~\ref{sec:Detectors} (Multi-Detector PGT-System) and \ref{sec:Electronics}  (Signal acquisition and waveform analysis). At sub-clinical rates, the CNAO ionization chamber-based beam monitors do not steer the beam. Therefore, we manually centred our system in the transverse plane by adjusting the proton detector and setup table to maximize the proton count rate through the LGAD strips.

Two target setups were used: a homogeneous PMMA phantom (30 cm long, with a transverse size of  $15\times15~\text{ cm}^2$, centred at the isocenter), and two PMMA phantom blocks (same transverse size and 15 cm of length) separated by a 4-cm gap. When using the latter, we had to radially shift back the detector in position 11 by 4~mm, to avoid collisions with the phantom during the measurement (see Figure \ref{fig:fig1a}).
A 226.9 MeV beam was directed at the homogeneous phantom for the stopping power measurements, while both target setups were used 
to assess the sensitivity of the technique to identify range shifts. The projected range in PMMA given by NIST PSTAR is 33.05 g/cm$^2$ , and the PMMA density quoted by the producer was 1.18 g/cm$^3$; hence, the expected beam range was 28.0 cm.

\subsubsection{System characterisation}
\label{sec:Characterization}
The performance of the detection system was characterised with a dedicated beam test. We used the homogeneous PMMA phantom and 226.9 MeV beam, as previously described. The two prompt-photon detectors were positioned upstream, at a distance of 20 cm from the isocenter, symmetrically with respect to the beam axis, and at angles of $\pm 50.7^\circ$. This position should lead to almost identical measurements (because of the symmetry), mainly composed of photons due to the upstream position.

\subsubsection{Signal acquisition and waveform analysis}
\label{sec:Electronics}

\begin{figure}[ht]
    \centering
    \includegraphics[width=0.8\linewidth]{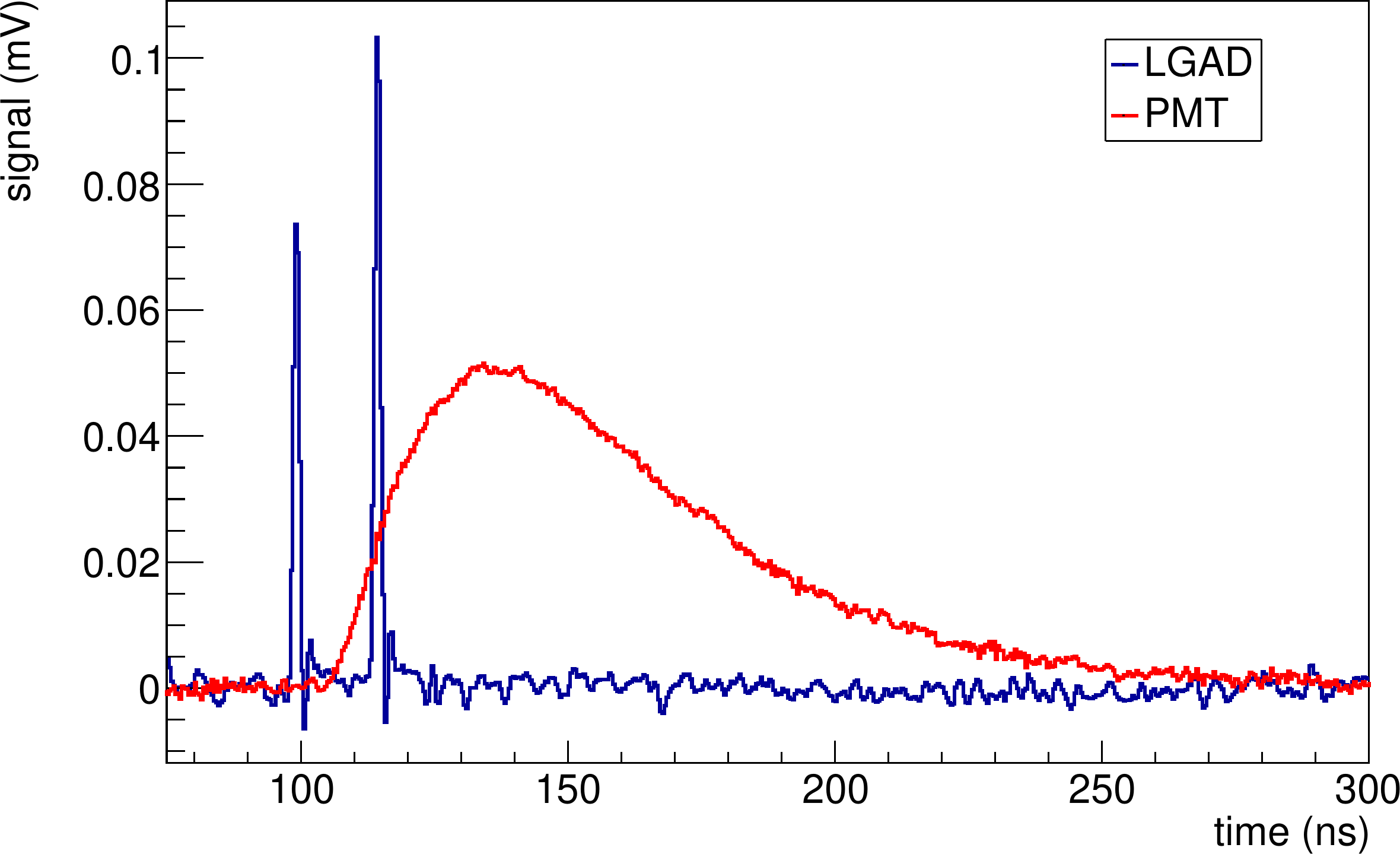}
    \caption{Waveform snapshots after baseline subtraction and signal inversion, showing the
photon signal in a PMT (red) and two proton signals in a LGAD-channel (blue).}
    \label{fig:fig2}
\end{figure}

The detector signals were acquired with a CAEN DT5742 waveform digitizer,  
which features two DRS4 ASICs of eight channels each. The digitizer was set up to sample the signals at 2.5~GHz, corresponding to a sampling time of 0.4~ns, with each waveform being 1024 samples long. The dead time between each signal acquisition was about 400 $\mu$s. We connected the two prompt-photon detectors, the reference detector, and strips 2-6 of the read-out amplifier board of the LGAD detector to the same DRS4 ASIC. The three channels connected to the photon detectors were configured to trigger the acquisition of the whole eight-channel block. The trigger threshold can be independently set for each detector, and we set them to levels equivalent to about 0.8~MeV, to avoid triggering on the 511-keV activation photons. An example waveform with one photon and two proton signals is shown in Figure~\ref{fig:fig2}.

To calculate the detection time of each particle, the acquired waveforms were stored and processed offline. Typically, software-based CFD techniques are implemented to correct the time-walk effect. However, in our case this was not possible because the signals were too coarsely sampled (in the case of the LGADs channels) or at risk of saturating the DAQ amplitude dynamic range (in the case of photodetector channels). Instead, the first step for all waveforms was baseline subtraction to account for possible fluctuations; next, the signals were identified using a threshold set above the noise level, and then discriminated according to their amplitude. Finally, the detection time was calculated by fitting the rising edge of the signal.

\subsection{Data analysis}
\subsubsection{Data processing and calibration}
The initial step in the timing analysis was to identify a photon-proton coincidence. Once a valid secondary signal was found  in a waveform snapshot, we computed the time difference between that secondary and all the protons in the snapshot (usually tens in number).
The time assigned to each coincidence had to be corrected for the delay of the corresponding LGAD strip and photon detector. For the photon detector, we use the measurement described in Section~\ref{sec:Electronics} (Signal acquisition and waveform analysis) to determine their relative delays. To obtain the delay between the photon detectors and the LGAD strips, the first configuration of the detectors was reproduced in a FLUKA simulation. The data were then analysed following Ferrero et al.\cite{Ferrero_2022a}, aligning the rising edges of the simulated and experimental PGT distributions to extract the relative delays.

The coincidences were stored into 50-bin histograms with 100 ps bins. Each detector and position were treated separately. This resulted in 14 histograms representing the PGT distributions measured by the two prompt-photon detectors in seven different positions, plus the 7 histograms of the reference detector, which was kept in the same position for all the measurements.
For the reconstruction of the emission distribution, the measurements performed with the 7 different detector configurations were treated as a single irradiation, i.e. as if they were measured with 14 detectors simultaneously. To ensure consistent measurements across all seven configurations, the prompt-photon detector coincidences were considered only until the reference detector reached a predefined number of photons ($4.0\cdot10^5$ for the homogeneous target, $4.7\cdot10^5$ for the cavity one).

\subsubsection{Emission reconstruction}
\label{sec:Recons}
The current implementation of SER-PGT, as described in Pennazio et al.\cite{Pennazio_2022}, relies on maximizing the Poisson likelihood via the Maximum Likelihood Expectation-Maximization algorithm.
For the present work, the algorithm was adapted to account for random coincidences in the PGT measurements.
\begin{equation}
\hat{x}_{jp}^{(k+1)} =
\frac{\hat{x}_{jp}^{(k)}}{\sum_{d} \sum_{n} h_{dnjp}}
\sum_{d} \sum_{n} 
\frac{h_{dnjp} y_{dn}}{b_{dn} + \sum_{j'} \sum_{p'} h_{dnj'p'} \hat{x}_{j'p'}^{(k)}}
\label{eq:1}
\end{equation}
where $\hat{x}_{jp}^{(k)}$ is the estimated number of prompt photons reconstructed at iteration $k$ emitted from spatial bin $j$ during the emission time bin $p$; $y_{dn}$ represents the measurement (i.e. the number of coincidences) from detector $d$ in the time detection bin $n$ ; $h_{dnjp}$ is an element of the system matrix, representing the probability of a photon emitted in $(j,p)$ being detected in $(d,n)$. Finally, $b_{dn}$ is the expected number of random coincidences in the measurement bin $(d,n)$. An estimate of $b_{dn}$ was obtained by analysing the PGT distribution with a Statistics-sensitive Non-linear Iterative Peak-clipping algorithm\cite{Morh__1997}, performed with 10 iterations. This method is usually applied to separate background and gamma peaks in (multi-dimensional) gamma ray spectra. In our case, it has proven to be effective in identifying the random coincidence background of PGT caused by the high number of protons passing through the LGAD detector in the same coincidence event.

\subsubsection{Model and calibration}

The system matrix in equation~(\ref{eq:1}) models the detection probability of prompt photons. 
As in our previous works\cite{Pennazio_2022,Ferrero_2022a,Werner_2024}, we calculated the system matrix elements using Monte Carlo simulations and considering the ideal case of identical detectors; this step provided the $h^{*}_{dnjp}$ elements for each detector $d$. The measurements performed with the two prompt-photon detectors in symmetrical positions (Section~\ref{sec:Characterization}) showed about 10\% difference in the number of detected coincidences. To account for this behaviour within the system matrix, we calculated an efficiency correction factor $c_d$ for each detector: $h_{dnjp} = c_d h^{*}_{dnjp}$.

\subsubsection{Range assessment}
\label{sec:Range}

The proposed method for range assessment depends on two hyperparameters, namely $k$ and $T$. 
The former is the number of MLEM iterations (~\ref{eq:1}) employed to reconstruct the spatiotemporal distribution of the emission, $\{\hat{x}_{jp}^{(k)}\}$. The latter is a threshold applied to $\{\hat{x}_{jp}^{(k)}\}$ to further remove background. 
For a given $(k,T)$ combination, a depth-versus-time profile $(t,z)$ is calculated as the average emission position within $\{\hat{x}_{jp}^{(k,T)}\}$ for each $p$. The range $R^{(k,T)}$ thus corresponds to the last spatial contribution of the depth-versus-time profile \cite{Pennazio_2022}.

In this study, we explored the features of the reconstructed  $\{\hat{x}_{jp}^{(k,T)}\}$ distributions to assess range deviations between two measurements. 
First, we studied the trend of the estimated range as a function of the MLEM iterations; therefore, we first determined $k_{\min}$, the minimum number of iterations  before which the range estimate significantly fluctuates. 
Then, we studied the distribution of the number of values in $\{\hat{x}_{jp}^{(k,T)}\}$ with respect to the threshold $T$; an acceptance window was set to prevent ineffective filtering (low $T$) or excessive erosion (high $T$). Next, $R^{(k,T)}$ was extracted from all the remaining 
$\{\hat{x}_{jp}^{(k>k_{\min},\,T_{\min}<T<T_{\max})}\}$ reconstructed distributions. 

Finally, we compared the reconstructed range $R^{(k,T)}$ of different subsets to investigate range shifts. Given two subsets $a$ and $b$, the difference $d^{(a-b,k,T)} = R^{(a,k,T)} - R^{(b,k,T)}$ was calculated for each $R^{(k,T)}$ pair as long as at least one of the two values of $T$ is in the required  interval, i.e., $T_{\min} < T < T_{\max}$. Then, a histogram was filled with all the $d^{(a-b,k,T)}$ values, and a 1-cm wide sliding window on the horizontal axis  identifies the most populated peak in the distribution. The mean of the values  within the interval identified by the window corresponds to the estimated range difference $d^{(a-b)}$.

\subsubsection{Stopping power measurement and comparison with expected value}
\label{sec:Power}

For the stopping power estimation, the ordered set of $(t,z)$ pairs calculated as in Section ~\ref{sec:Range} serves as a stand-in for the unknown primary particle motion $z=z(t)$, since the PG-emission is closely linked to the presence of primary particles in time and space \cite{Ferrero_2022b}. The $(t,z)$ pairs were fitted to the motion model proposed in Werner et al.\cite{Werner_2024}, obtaining an analytical estimate of the average proton motion. 
The stopping power was then calculated using $S = -\frac{dE}{dz} = \frac{1}{p \sqrt{\alpha}} (R_0 - z)^{\tfrac{1}{p}-1}$, where $\alpha$ and $p$ are the free model parameters; $R_0 = \alpha E_0^p$ is the particle range for initial energy $E_0$ and $z$ is the depth within the material \cite{Bortfeld_1996}, respectively. The model assumes the material to be homogeneous.

The optimum number of iterations $k$ and threshold $T$ were found using a risk-sensitive approach \cite {Werner_2024}. The optimization minimizes the mean and standard deviation of the Mean Relative Error (MRE) of the estimated stopping power profile. We considered up to 4000 iterations and thresholds relative to the maximum of the distribution from 0 to 0.8 in steps of 0.025. This data-driven optimization requires the data to be separated into training and evaluation sets. The estimated stopping power was evaluated using the MRE of the whole stopping power profile and the Stopping Power Ratio (SPR) to water at 100 MeV \cite{Werner_2024}. The NIST PSTAR database served as the reference.

\section{Results}
\subsection{Data acquisition and statistics}

The proton detector acquisition was triggered by the secondary particle detectors; hence, it could not provide unbiased proton rate measurements during the PGT measurement. Therefore, the primary particle rate was measured in advance. This measurement, corrected by the detector efficiency, indicated about  $2.9~ - ~ 6.6\cdot 10^7$ irradiated protons with an average rate of about  $2~ - ~ 4\cdot 10^7$ pps.

During the irradiation of the homogeneous phantom, the reference detector collected $4.0\cdot 10^5$ events for each position of the two prompt-photon detectors, while it detected $4.7\cdot 10^7$ events during the irradiation of the phantom with the air gap. The number of events detected by the reference detector was used to divide each of the seven-position acquisitions into ten equally populated subsets, to be reconstructed separately. The number of particles in each subset approximately corresponds to the particle statistics typically found in a highly populated spot of a proton therapy treatment delivered with pencil beam scanning.

\begin{figure}[ht]
    \centering
    \begin{subfigure}{0.48\linewidth}
        \centering
        \includegraphics[width=\linewidth]{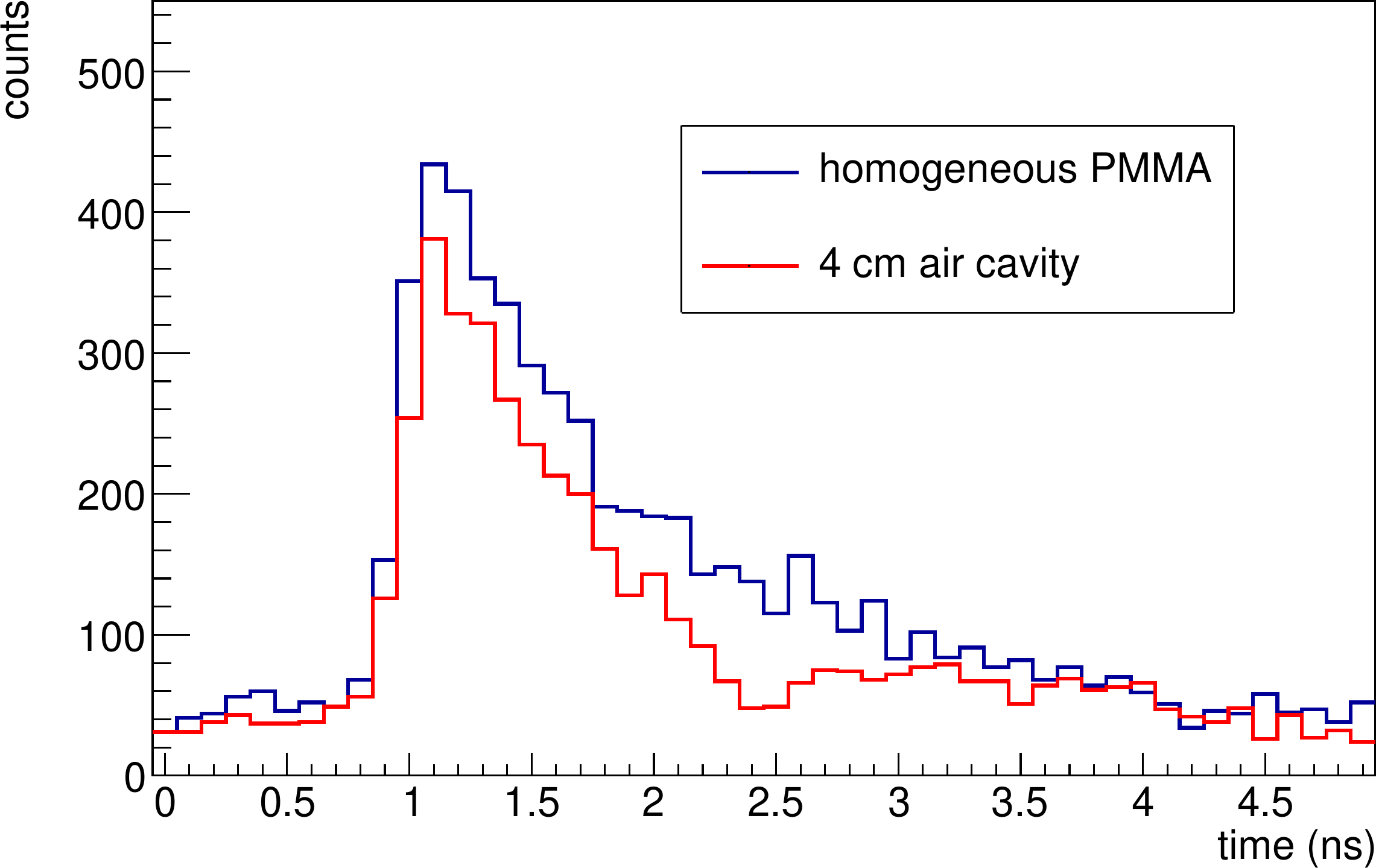}
        \caption{}
        \label{fig:fig3a}
    \end{subfigure}
    \hfill
    \begin{subfigure}{0.48\linewidth}
        \centering
        \includegraphics[width=\linewidth]{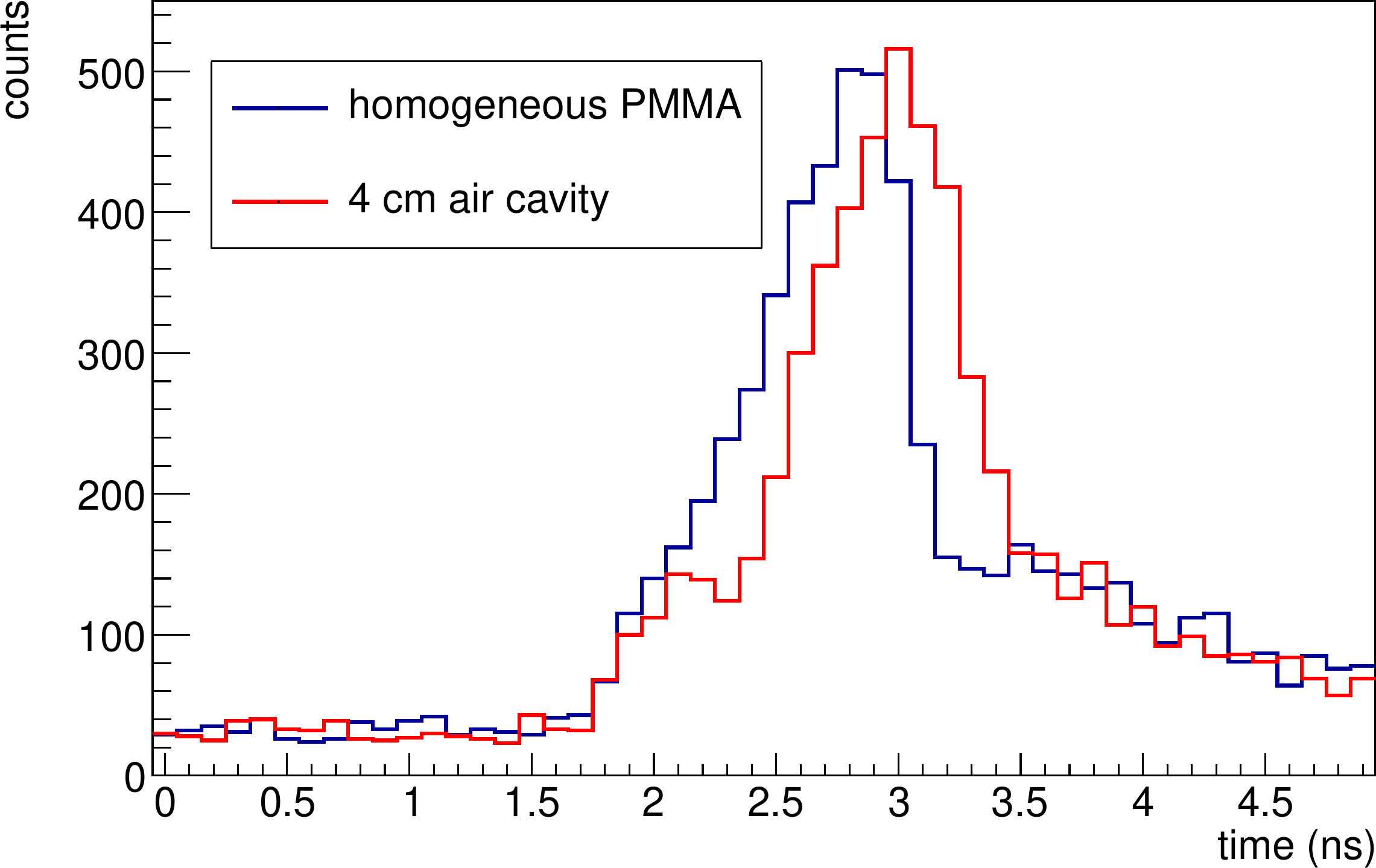}
        \caption{}
        \label{fig:fig3b}
    \end{subfigure}
    \caption{Example of PGT spectra of the first subset with and without the air cavity for detector
positions 1 (a) and 11 (b).}
    \label{fig:fig3}
\end{figure}

The estimation of the random coincidences was performed as described in Section~\ref{sec:Recons}, exploiting the setup with the homogeneous phantom and scaling down the resulting distribution by 1/10, to adapt it to the subsets size. We used the same random distribution to reconstruct
the data of the phantom with the air cavity.
The selected time distributions in Figure~\ref{fig:fig3} show the homogeneous phantom and air-gap measurements. As expected, the spectra are different in shape and statistics because of different target/detector positions.

\begin{figure}[ht]
    \centering
    \includegraphics[width=\linewidth]{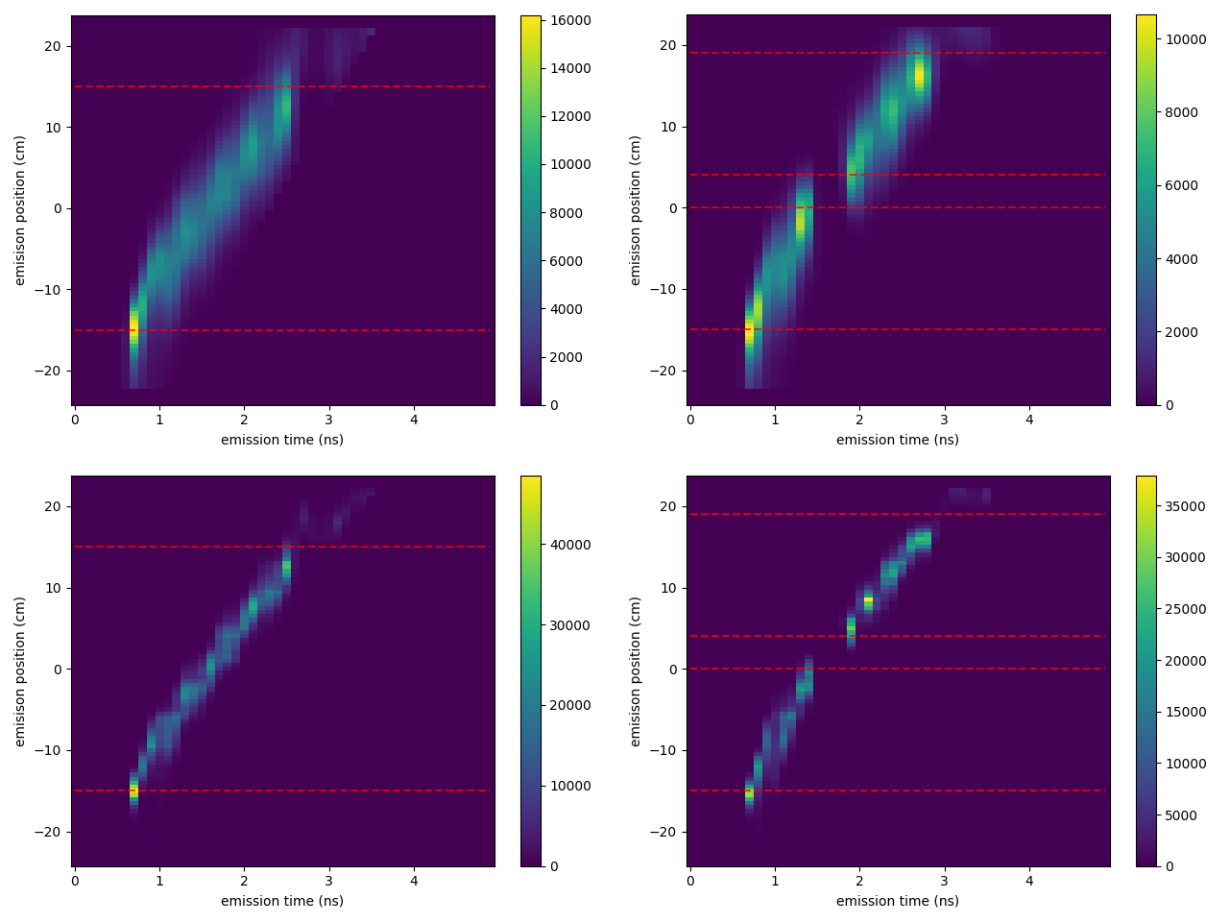}
    \caption{Reconstructed spatiotemporal distributions using the first subset of PGT-data. Left:
homogeneous phantom after 100 (top) and 1000 (bottom) iterations. Right: 4-cm air cavity
after 100 (top) and 1000 iterations (bottom). No threshold was applied to these distributions.
Spatial phantom boundaries are marked with red dashed lines.}
    \label{fig:fig4}
\end{figure}

The reconstructed spatiotemporal distributions shown in Figure~\ref{fig:fig4} provide information about the underlying motion of primary particles. High intensity values are found at the phantom entrance, probably a consequence of the Gibbs phenomenon \cite{Tong_2011}. More iterations reduced the spatial extension of the distribution but increased intensity oscillations. While the resulting artifacts may hinder stopping power estimation at high iterations, the improved resolution in the spatial dimension is advantageous for range estimation. In both cases, the air gap, which causes a shift in the distribution, is clearly visible, especially in the temporal dimension.

\subsection{Range difference measurement}
The reconstruction algorithm in equation~(\ref{eq:1}) was applied up to $k = 4000$ iterations, for which $R^{(k,T)}$, averaged over $T$, changed less than 1 mm over the last 500 iterations. We used the first subset of the homogeneous target acquisition, setting $k_{min}  = 100$. $T$ was selected to ensure that  $\{\hat{x}_{jp}^{(k,T)}\}$ contains at least 50 nonzero elements
when $T = T_{min}$, but less than 500 nonzero elements when $T = T_{max}$. The same condition was applied for all the remaining datasets of both homogeneous and cavity phantoms.

\begin{figure}[ht]
    \centering
    \includegraphics[width=0.8\linewidth]{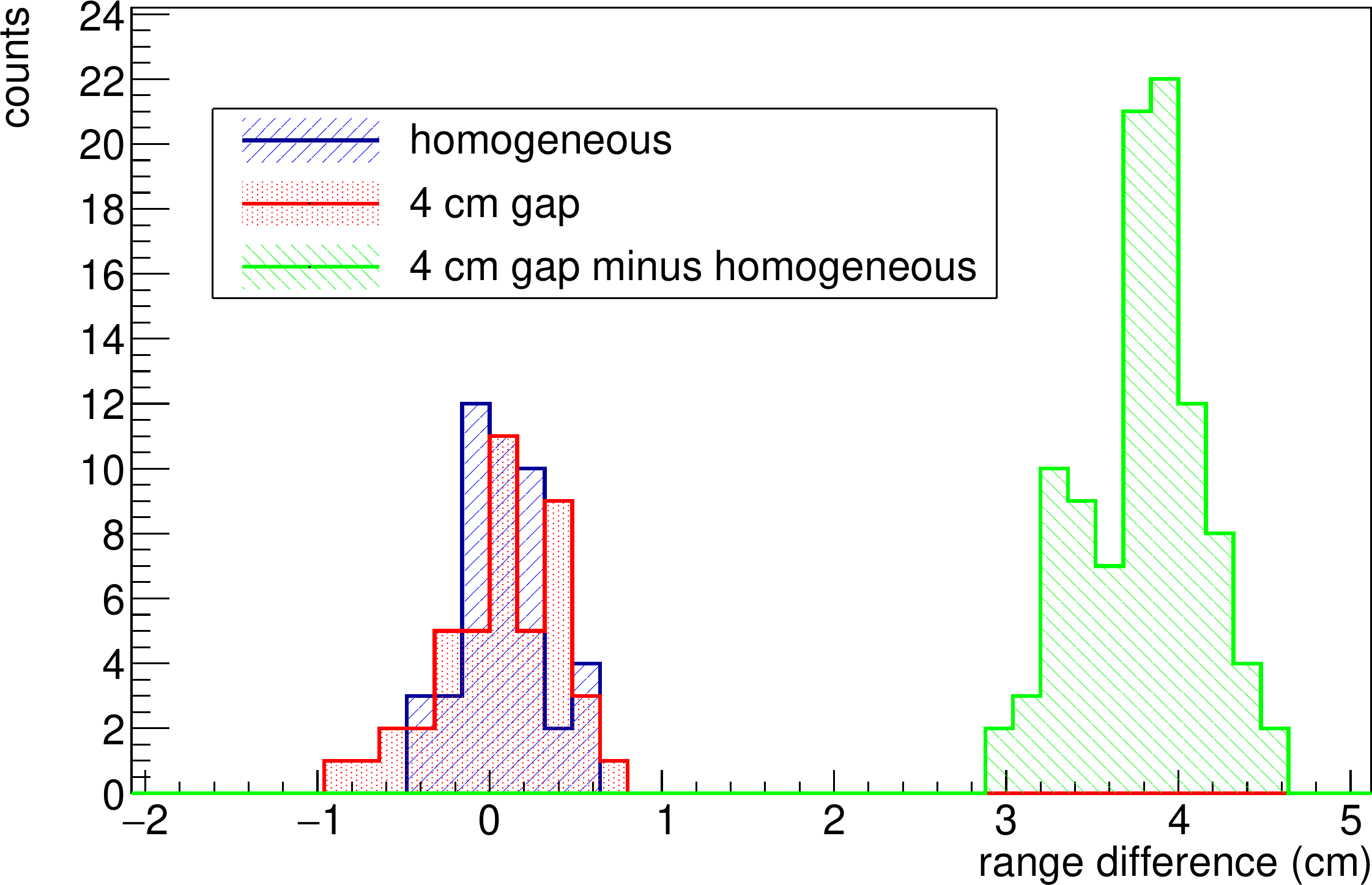}
    \caption{Range differences for homogeneous subsets (blue) and 4-cm air gap subsets (red), both centred at 0.08cm (± 0.3 cm). Differences between the air-gap and homogeneous subsets (green) show the expected 3.8 cm shift (± 0.3 cm).}
    \label{fig:fig5}
\end{figure}

We calculated $d^{(a-b)}$ by comparing the subsets of the homogeneous target with those of the target containing the 4 cm air cavity. With 10 subsets per configuration, this yielded 100 possible comparisons. The resulting range difference was 3.8 cm (± 0.3 cm, one standard deviation). We also computed $d^{(a-b)}$ using  subsets from the same configuration (i.e. 49 comparisons for both the homogeneous and cavity phantoms), obtaining a range difference of 0.08 cm (± 0.3 cm). The histograms in Figure~\ref{fig:fig5} show the internal consistency of subsets belonging to the same geometry, as well as the expected 4 cm difference between the homogeneous and air-gap measurements, neglecting the air density.

\subsection{Stopping power measurement}
\begin{figure}[ht]
    \centering
    \begin{subfigure}{0.48\linewidth}
        \centering
        \includegraphics[width=\linewidth]{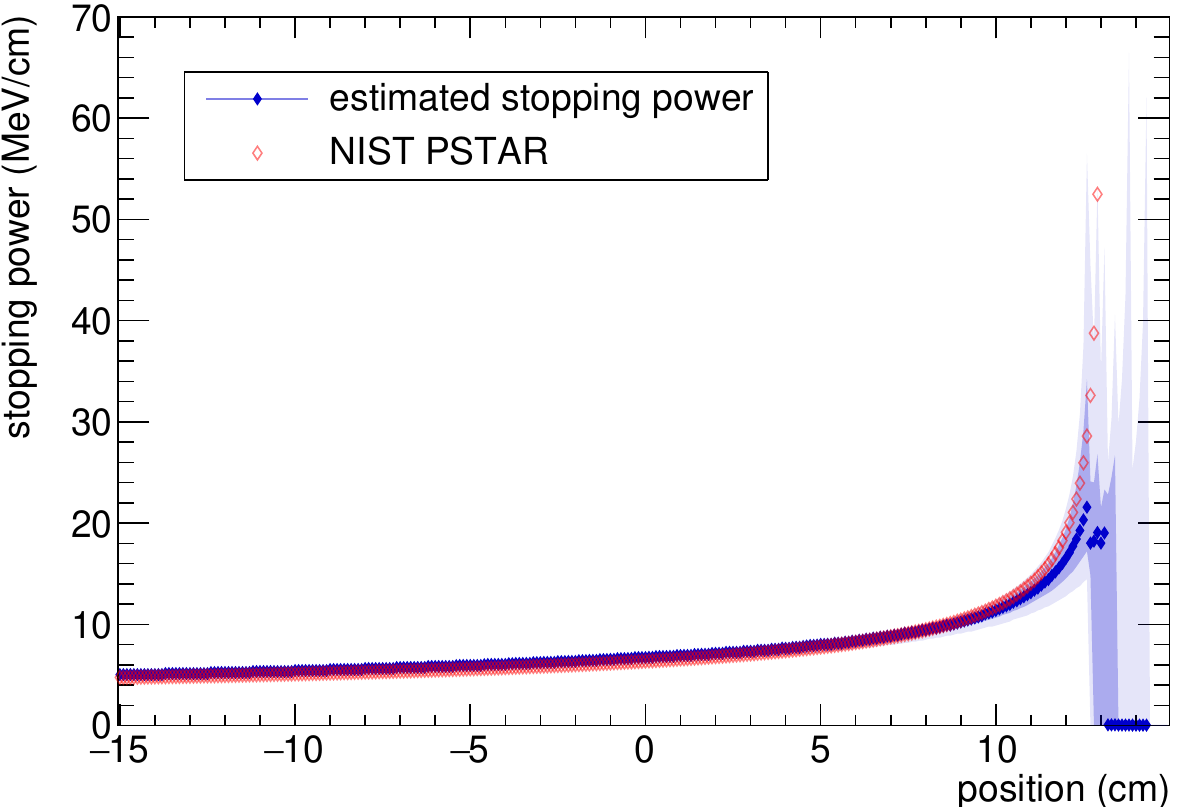}
        \caption{}
        \label{fig:fig6a}
    \end{subfigure}
    \hfill
    \begin{subfigure}{0.48\linewidth}
        \centering
        \includegraphics[width=\linewidth]{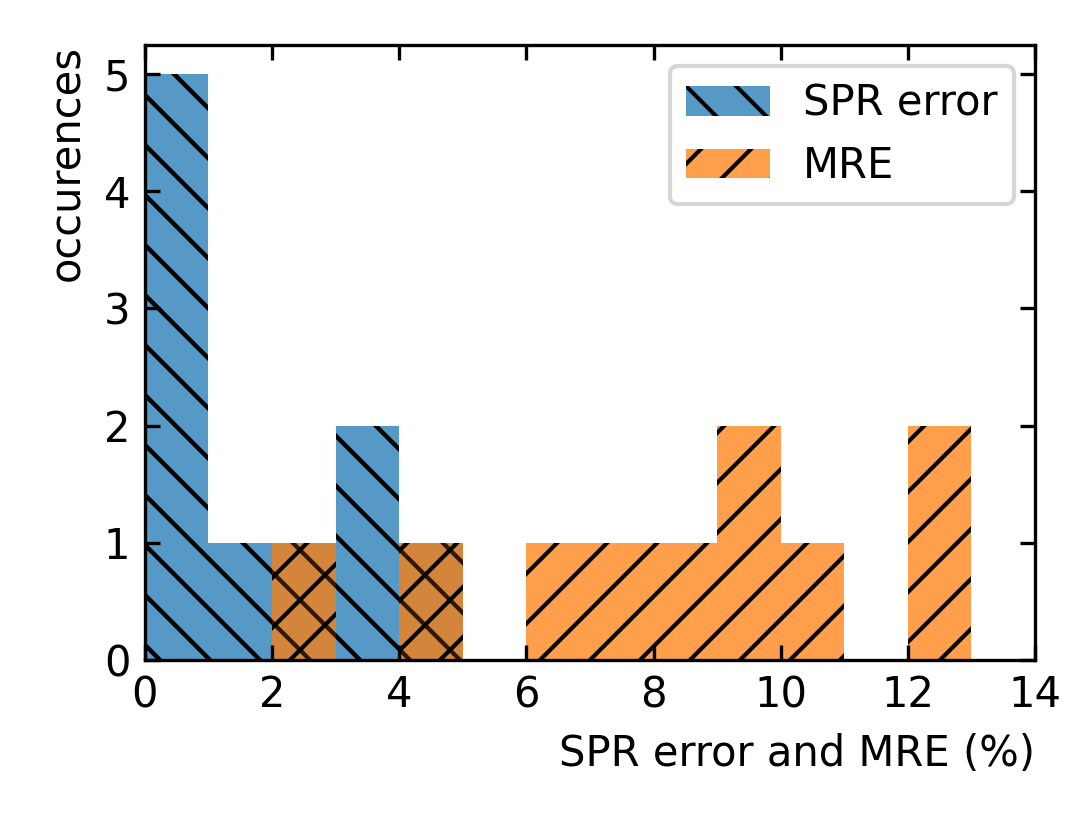}
        \caption{}
        \label{fig:fig6b}
    \end{subfigure}
    \caption{(a): Stopping power versus depth. The reference values are given in red, while the
blue diamonds represent the median of the ten estimated stopping power curves for the
homogeneous phantom; the corresponding quartiles are described by the boundaries of
shaded regions. (b): histograms of the SPR at 100 MeV and the MRE of the stopping power
curves.}
    \label{fig:fig6}
\end{figure}

Since ten datasets were available, we employed the leave-one-out technique: Each dataset was evaluated using the parameters optimized based on the remaining nine datasets. The stopping power was only derived from the measurements without the air gap, since the
technique has  not yet been extended to non-homogeneous phantoms.
Figure~\ref{fig:fig6} compares the estimated and the reference stopping power from NIST PSTAR. The MRE of the estimated stopping power profile, averaged over all depth bins and data subsets, was 8\% ± 3\%. The error of the SPR was 2\% ± 2\%. Note that only the MRE was explicitly minimized using the training set. The number of iterations $k$ chosen by the optimization procedure was between 39 and 46 iterations, even though iterations up to 4000 were considered. The applied relative thresholds $T$ were between 0.325 and 0.500.

\section{Discussion}
The results shown here represent the experimental validation of several novel techniques. We reported the first PGT-measurements using a synchrotron proton beam and using an LGAD as the primary particle detector. Moreover, we carried out the first spatiotemporal reconstructions with SER-PGT based on experimental data, as well as the subsequent estimations of the range and the stopping power. 
These results also confirmed that prompt-photon emission reflects the average proton motion inside the target.
The non-invasive stopping power measurement using a therapeutic beam and based on primary particle kinematics is especially significant.
It is the only technique published so far that potentially allows on-the-fly measurement of the stopping power in proton therapy. Other off-line approaches require higher energy beams (proton CT) or indirect measurements and calibration against phantoms (CT and dual-energy CT).

The 2D spatiotemporal distribution from SER-PGT allows a direct estimation of the kinetic energy loss by the beam, which is the actual definition of the total stopping power. For proton beams above 1 MeV, the contribution of the nuclear stopping power is less than 0.1\% (https://physics.nist.gov). Therefore, for most proton beam energies in particle therapy, the total stopping power is a good approximation of the electronic stopping power, which, by definition, corresponds to the unrestricted LET. However, it is important to note that further Monte Carlo-based studies are required to accurately define the correspondence between the stopping power measured through SER-PGT and the variously defined LET averaging methods \cite{Kalholm_2021}.
In-vivo verification of the stopping power is of potential interest for LET-guided treatment plan optimisation, which aims to improve the radiobiological effectiveness \cite{McIntyre_2023}. LET optimization in particle therapy is experiencing a renaissance, driven by the technological advancements and increasingly new particle therapy centres \cite{Molinelli_2024}.

The PGT-spectra shown in Figure~\ref{fig:fig3} demonstrate the feasibility of PGT-measurements in a synchrotron-based facility, although at a sub-clinical rate. Improvements to the readout electronics are needed to enable effective simultaneous measurements with more channels and to handle the high instantaneous particle rates of synchrotrons.
The spatiotemporal distributions shown in Figure~\ref{fig:fig4} are the first SER-PGT results based on experimental data. At a lower number of iterations ($k < 100$), the intensity along the proton path is stable, although blurred in the spatial dimension. At higher iterations artefacts akin to the Gibbs phenomenon appear, as hinted at in Werner et al.\cite{Werner_2024}. While the resolution for edges and range greatly improves with more iterations, the artefacts hinder an accurate estimate of the motion. However, when choosing the reconstruction parameters carefully, the primary particle motion can be extracted from multi-detector PGT sets. 

The expected number of random events was determined from the full PGT measurement of the homogeneous phantom and used for reconstructing the subsets with and without the air cavity in the phantom. If available, a custom estimate for each dataset may improve the reconstruction results further. Furthermore, no attenuation or scatter corrections were applied in the reconstruction step. The impact of attenuation and scatter was negligible in Monte Carlo simulations\cite{Pennazio_2022}. Implementing corrections for scatter and attenuation requires prior information about the target composition, which may introduce bias when using this technique for therapy verification.

We plan refinements of SER-PGT to further improve the results and move closer to clinical relevance. Even though the current background model in the PGT-spectra worked adequately within this study, its shape still needs better characterization. The events contributing to it includes slower secondary particles and random proton-photon coincidences. Estimating the random components requires accurate knowledge of the fine time structure of the delivered beam, which is not yet available.  
Figure~\ref{fig:fig5} shows range difference estimations with a 0.3~cm standard deviation, consistent with previous simulation-based results \cite{Ferrero_2022b, Werner_2024}. The range difference of 4~cm is recovered with a slight bias (underestimation of 0.2~cm). A range verification system needs to be able to resolve range shifts on the order of millimetres. Here we demonstrate that the 4-cm air gap is within the resolution capabilities of our system, motivating further measurements of smaller range differences. 
The stopping power estimation reported in Figure~\ref{fig:fig6} is consistent with the simulation-based results for 220-MeV protons of Werner et al.\cite{Werner_2024}, even though only 14 detectors were used here, compared to the 110 detectors of that work.
Further measurements with low beam energies are needed to better validate the results of Werner et al.\cite{Werner_2024}. 

Several issues remain before assessing clinical viability. The current model to estimate the stopping power assumes a homogeneous phantom. The extension to include heterogeneous targets is currently underway.
The sensitive area of the LGAD detector needs to be enlarged to allow for lateral beam scanning. This will also increase the number of detected coincidences, as the size of the detector used in this study was smaller than the beam transverse size. By increasing the number of photon detectors, a dedicated study could be carried out to identify the most informative detector geometry and, consequently, the minimum number of primary particles needed to extract meaningful stopping power and range information.

The instrumentation required for the SER-PGT technique is less complex compared to other proton therapy monitoring modalities, such as in-beam PET and gamma cameras, still, the proposed method is an important step toward in-vivo stopping power measurements. Indeed, SER-PGT followed by the stopping power retrieval procedure demonstrates considerable potential, thereby supporting further technical refinement and methodological development.

\bigskip

\bibliographystyle{ieeetr}
\bibliography{references1}

\section*{Acknowledgements}
This work is partially supported by the German Research Foundation (DFG) by Grant No. 516587313 (PROSIT).

\newpage 

\end{document}